# Constraints on the energy spectra of charged particles predicted in some model interactions of hadrons with help of the atmospheric muon flux

L G Dedenko[1,2], G F Fedorova[1], T M Roganova[1]

[1]Skobeltsyn Institute of Nuclear Physics, Lomonosov Moscow State University, 119991 Moscow, Russia

[2]Faculty of Physics, Lomonosov Moscow State University, 119991 Moscow, Russia

E-mail: ddn@dec1.sinp.msu.ru

**Abstract**
It has been shown that muon flux intensities calculated in terms of the EPOS LHC and EPOS 1.99 models at the energy of $10^4$ GeV exceed the data of the classical experiments L3+Cosmic, MACRO and LVD on the spectra of atmospheric muons by a factor of 1.9 and below these data at the same energy by a factor of 1.8 in case of the QGSJET II-03 model. It has been concluded that these tested models overestimate (underestimate in case of QGSJET II-03 model) the production of secondary particles with the highest energies in interactions of hadrons by a factor of ~1.5. The LHCf and TOTEM accelerator experiments show also this type of disagreements with these model predictions at highest energies of secondary particles.

## 1. Introduction

The extensive air showers (EAS) are used to understand the origin and the composition of cosmic rays, their possible sources and a transport of cosmic particles in various magnetic fields on their way to the Earth at very high energies. All features of the energy spectrum, arrival directions and a composition of the primary cosmic ray particles should be determined through the analysis of the EAS data. These data as some signals in the surface and underground detectors are usually interpreted in terms of some models of hadronic interactions [1−6]. Usually these models are tested with the help of the accelerator data at small values of the pseudorapidity $\eta$ ~0 where most of secondary particles are produced [7–9]. However, calculations show that the energy flow of secondary particles reaches its maximum at values of the pseudorapidity $\eta$ in the interval of 6−10 at energy 14 TeV in the CMS [10]. Let us also note that the longitudinal development of EAS and, hence, the depth $X_{\max}$ of its maximum depends strongly on the rate of the projectile particle energy fragmentation. If a probability of secondary particle production in the energy range near the energy of the projectile particle is high then the depth $X_{\max}$ is expected to be rather large. And contrary, in case of the severe energy fragmentation the length of a shower and the depth $X_{\max}$ of its maximum will be small. The intensity of the muon flux in the atmosphere depends also on the number of secondary particles produced with the maximal possible energies. So, this energy interval is the most important for the EAS longitudinal development. The study of the secondary particle production with the most highest energies is also of importance for understanding of hadronic interactions. Some models [2−6] at the most highest energies of secondary particles are tested at the accelerator experiments LHCf [11] and TOTEM [12].

As an example, let us note that in the standard approach the depth $X_{max}$ of shower maximum as a function of the energy $E$ of the primary particles is used to study a composition. For various rates of this energy fragmentation the depths $X_{max}$ and, hence, a predicted composition will be different. In the alternative approach the ratio $a$ of signals $s_\mu(600)$ in the underground and $s(600)$ in the surface detectors at a distance of 600 m from the shower core is used to study the nature of the primary particles:

$$\alpha = s_\mu(600)/s(600). \qquad (1)$$

The Yakutsk array data [13] interpreted in terms of the model QGSJET II-03 [2] predicted the heavy composition [14]. But even with the rude correction of a fragmentation rate this model showed the light composition at energies $10^9$−$10^{10}$ GeV [15]. It is well known that in the atmosphere $\mu^\pm$ mesons are mainly generated through decays of $\pi^\pm$ and $K^\pm$ mesons produced in cascades induced by the various primary cosmic ray particles with different energies. The energy spectra $dI_j/dE$ of the primary particles of type $j$ may be well approximated in some energy intervals by the power low:

$$dI_j/dE = A_j \cdot E^{-\gamma}, \qquad (2)$$

where the exponent $\gamma = 2.7$−$3.1$ depends on the energy $E$. As these spectra are very steep it is evident that only $\pi^\pm$ and $K^\pm$ mesons produced with maximal possible energies in interactions of hadrons contribute mainly to the atmospheric muon flux. Due to steepness of the primary particle energy spectra such high energetic $\pi^\pm$ and $K^\pm$ mesons are substantially produced in cascades induced by the primary protons and helium nuclei as the energy per nucleon is of importance. So, the study of the atmospheric muon flux may clarify the mechanisms of $\pi^\pm$ and $K^\pm$ mesons production with maximal energies near the energy of the projectile particle. One of goals of this paper is an estimation of the energy interval where this contribution would be most considerable.

As the atmospheric muon flux depends strongly on the rate of the projectily particle energy fragmentation we suggest to test the various models of hadronic interactions by their predictions of this flux. It should be noted that this test is the most sensitive one to a production of secondary charges particles with the maximal possible energies. In fact, in papers [16−19] some low energy models and the package FLUKA [20] have been tested in such a way. The atmospheric muon energy spectra have been calculated in terms of some models at energies above $10^2$ GeV by solving the transport equations [21].

We suggested [22−24] the original method of simulation of the atmospheric muon energy spectrum with the help of the package CORSIKA [25] to test also the most popular models of hadronic interactions with the atmospheric muon data [26−28], measured with rather high accuracy at energies above $10^2$ GeV. This test is of great importance for the study of a composition and the energy spectrum of the primary particles at high energies. In our paper [24] we have tested the QGSJET 01 [1], QGSJET II-04 [3] and SIBYLL 2.1 [4] models and have found out that these models predict more intense muon flux than data [26−28] by factors of 1.7−2 an energy $10^4$ GeV of muons. In the conference papers [22, 23] we have preliminary tested models QGSJET II-03 [2] and EPOS 1.99 [5].

In this paper our original method of simulation of the atmospheric muon spectrum have been used to test the models EPOS LHC [6] and in detail QGSJETII-03 [2] and EPOS 1.99 [5] with the help of the smooth approximation of the atmospheric muon data observed by the collaborations L3+Cosmic [26], MACRO [27] and LVD [28]. A comparison of these muon data with our results of simulations of the muon energy spectrum allows to draw a conclusion about secondary particles production with the maximal energies.

**2. Simulations**

The atmospheric muon energy spectrum may be calculated by various approaches. The transport equations describing a propagation of different particles in the atmosphere may be solved to find out this spectrum [21, 29−31]. The Monte Carlo method allows to get some estimate of this muon spectrum [32−35]. We had suggested a very simple original variant [14, 22−24] of the Monte Carlo approach. For the primary particle of type $j$ with the fixed energy $E$ the EAS are simulated with the help of the package CORSIKA [25]. No thining option has been used. All muons at the level of observation in these showers may be distributed in some energy histogram. If statistics of simulated showers is high enough this histogram will approximate rather well the distribution $S_j(E_\mu, E)dE_\mu$ of muons on the energy $E_\mu$ for the given primary particle of type $j$ with the fixed energy $E$. These simulations should be repeated for the various primary particles with different energies. As the number of the primary particles of the type $j$ with the given energy $E$ is determined by the energy spectrum $dI_j/dE$ of these particles it is straightforward to estimate the energy spectrum $D(E_\mu)dE_\mu$ of atmospheric muons as follows. It is evident that distribution $S_j(E_\mu, E)dE_\mu$ multiplied by the weight function $dI_j/dE$ and integrated on the energy $E$ is a contribution to the muon spectrum by the primary particles of the type $j$. The suggested method uses the energy spectra $dI_j/dE$ of the primary particles of all types $j$ and all the distributions $S_j(E_\mu, E)dE_\mu$ of muons on energy $E_\mu$ produced at the level of observation in showers induced by the primary particles of all types $j$ with the various fixed energies $E$.

So, the energy spectrum of the atmospheric muons $D(E_\mu)dE_\mu$ should be calculated as follows:

$$D(E_\mu)dE_\mu = \sum_j \int dE \cdot (dI_j/dE) \cdot S_j(E_\mu, E) \cdot dE_\mu \qquad (3)$$

Here we have to integrate on energy $E$ of the primary particles and to sum on all types $j$ of these particles. The energy spectra $dI_j/dE$ of the primary particles are very steep. As the energy spectrum per nucleon is of importance for the muon production then only the primary protons ($j=p$) and helium nuclei ($j=He$) should be taken into account.

Thus, our original method (3) had been used to estimate in terms of the QGSJET II-03 [2], EPOS 1.99 [5] and EPOS LHC [6] models the energy spectra $D(E_\mu)dE_\mu$ of the vertical atmospheric muons in the energy interval of $10^2-10^5$ GeV. This interval had been devided into 60 equal bins in the logarithmic scale (the width of a bin in this scale is equal to $h = 0.05$). The average energy $\langle E_\mu(i) \rangle$ of muons in the bin with the number $i$ was determined as $\langle E_\mu(i) \rangle = 10^{2+h(i-0.5)}$ where $i = 1, 2,...60$. For example, the average energies of muons for the bins with numbers 1, 21 and 41 which we will use later as some illustrations of simulations are equal to $1.059 \times 10^2$, $1.059 \times 10^3$ and $1.059 \times 10^4$ GeV, respectively.

So, we need to know the primary particle energy spectra $dI_p/dE$ and $dI_{He}/dE$. At energies $E \leq E_1 = 3 \times 10^6$ GeV we have used Gaisser T. K. and Honda M. (GH) approximation [36] of the energy spectra for the primary protons and helium nuclei:

$$dN_A/dE_k = K \cdot (E_k + b \cdot \exp(c\sqrt{E_k}))^\alpha, \qquad (4)$$

where the parameters *a*, *K*, *b* and *c* are assumed to be as 2.74, 14900, 2.15 and 0.21 for the primary protons (A = 1) and as 2.64, 600, 1.25 and 0.14 for the primary helium nuclei (A = 4). As an energy $E$ is always above $10^2$ GeV for the primary protons (and above $4\times10^2$ GeV for the helium nuclei) the kinetic energy $E_k$ is practically equal to $E$. We will denote GH approximations of the energy spectra as $(dI_p/dE)_{GH}$ and $(dI_{He}/dE)_{GH}$. We suggested the modified GH approximations of the energy spectra $(dI_j/dE)_m$ of the primary particles at energies $E > E_1$. For the primary protons and helium nuclei they look as follows:

$$(dI_p/dE)_m = (dI_p/dE)_{GH} \cdot (E_1/E)^{0.5} \tag{5}$$

and

$$(dI_{He}/dE)_m = (dI_{He}/dE)_{GH} \cdot (E_1/E)^{0.5}, \tag{6}$$

where $E_1 = 3\times10^6$ GeV.

Figure 1 shows the energy spectra of the primary protons and helium nuclei. At this figure the GH [36] at energies below $E_1$ and the modified GH approximations (5) and (6) at energies above $E_1$ are shown as dashed line. The various data [37−40] are shown by different symbols deciphered in captions to figure.

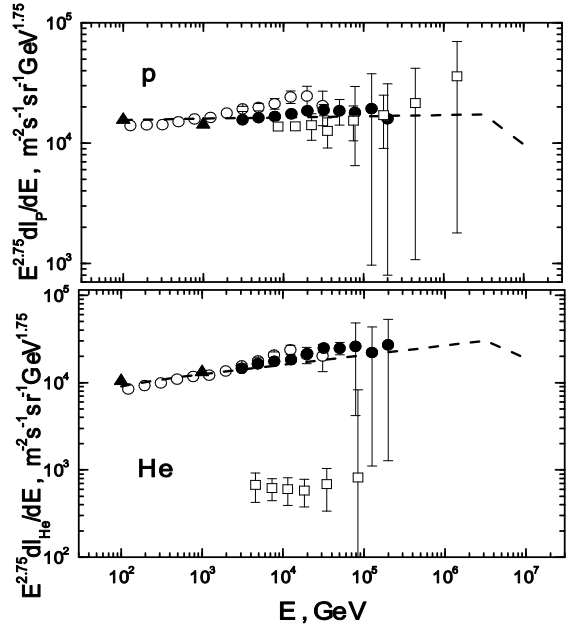

**Figure 1**. The energy spectra of the primary protons and helium nuclei. Dashed line − the GH [36] and the modified GH approximations; ○ − ATIC-2 [37], ● − CREAM [38], □ − RUNJOB [39], ▲ − AMS02 [40].

To estimate integrals (3) we have to calculate the distributions of muons $S_p(E_\mu,E)dE_\mu$ and $S_{He}(E_\mu,E)dE_\mu$ on energy $E_\mu$ in showers induced by the primary protons and helium nuclei with various fixed energies $E$ in terms of various models with the help of the package CORSIKA [25]. Calculations have been carried out for 24 values of energy $E$ for the primary protons and for 19 values of energy $E$ for the primary helium neclei. Statistics of simulated events are shown in table 1.

**Table 1**. Statistics of events for various energies E of the primary protons and helium nuclei.

| E, TeV | P | | | He | | |
|---|---|---|---|---|---|---|
| | QGSJII-03 | EPOS 1.99 | EPOS LHC | QGSJII-03 | EPOS 1.99 | EPOS LHC |
| 0.1333 | $10^6$ | $10^6$ | $10^6$ | | | |
| 0.1778 | $10^6$ | $10^6$ | $10^6$ | | | |
| 0.2371 | $10^6$ | $10^6$ | $10^6$ | | | |
| 0.3162 | $10^6$ | $10^6$ | $10^6$ | | | |
| 0.4217 | $10^6$ | $10^6$ | $10^6$ | | | |
| 0.5623 | $10^6$ | $10^6$ | $5\times10^5$ | $10^6$ | $10^5$ | $5\times10^5$ |
| 0.7498 | $10^6$ | $10^6$ | $5\times10^5$ | $10^6$ | $10^5$ | $5\times10^5$ |
| 1 | $10^6$ | $5\times10^5$ | $5\times10^5$ | $10^5$ | $10^5$ | $5\times10^5$ |
| 1.778 | $5\times10^5$ | $5\times10^5$ | $5\times10^5$ | $10^5$ | $10^5$ | $5\times10^5$ |
| 3.162 | $2.5\times10^5$ | $2.5\times10^5$ | $2.5\times10^5$ | $10^5$ | $10^5$ | $2.5\times10^5$ |
| 5.623 | $2.5\times10^5$ | $2.5\times10^5$ | $2.5\times10^5$ | $10^5$ | $10^5$ | $2.5\times10^5$ |
| 10 | $10^5$ | $10^5$ | $10^5$ | $5\times10^4$ | $10^5$ | $10^5$ |
| 17.78 | $10^5$ | $10^5$ | $10^5$ | $5\times10^4$ | $10^5$ | $10^5$ |
| 31.62 | $10^5$ | $5\times10^4$ | $5\times10^4$ | $5\times10^4$ | $10^5$ | $5\times10^4$ |
| 56.23 | $5\times10^4$ | $5\times10^4$ | $5\times10^4$ | $5\times10^4$ | $10^5$ | $5\times10^4$ |
| 100 | $5\times10^4$ | $10^4$ | $5\times10^4$ | $10^4$ | $10^4$ | $10^5$ |
| 177.8 | $10^4$ | $10^4$ | $3\times10^4$ | $10^4$ | $10^4$ | $10^5$ |
| 316.2 | $10^4$ | $5\times10^3$ | $10^4$ | $10^4$ | $10^4$ | $5\times10^4$ |
| 562.3 | $10^4$ | $10^4$ | $10^4$ | $10^4$ | $10^4$ | $5\times10^4$ |
| 1000 | $5\times10^3$ | $10^3$ | $4\times10^3$ | $5\times10^3$ | $10^3$ | $4\times10^3$ |
| 1778 | $5\times10^3$ | $10^3$ | $4\times10^3$ | $5\times10^3$ | $10^3$ | $4\times10^3$ |
| 3162 | $5\times10^3$ | $5\times10^2$ | $2\times10^3$ | $5\times10^3$ | $10^3$ | $2\times10^3$ |
| 5623 | $5\times10^3$ | $5\times10^2$ | $10^3$ | $5\times10^3$ | $10^3$ | $2\times10^3$ |
| 10000 | $10^2$ | $10^2$ | $3\times10^2$ | $10^2$ | $10^2$ | $4\times10^2$ |

### 3. Results and conclusions

The distributions $S_p(E_\mu, E)dE_\mu$ and $S_{He}(E_\mu, E)dE_\mu$ of the muons on energy $E_\mu$ calculated in terms of the EPOS LHC model for some fixed values of the energy $E$ of the primary protons and helium nuclei are presented in figure 2. The various curves are marked by digits which denote some fixed values of the energy $E$, shown in caption to figure. The densities $S_p(E_\mu, E)$ and $S_{He}(E_\mu, E)$ of these calculated distributions are decreasing with growth of the variable $E_\mu$. In small intervals of the energy $E_\mu$ this decreasing may be approximated by the power law:

$$S_p(E_\mu, E) \sim E_\mu^{-\alpha}, \qquad S_{He}(E_\mu, E) \sim E_\mu^{-\alpha}, \tag{7}$$

where $\alpha$ is increasing with growth of the energy $E_\mu$. The left part of curves at high energies of the primary particles may be approximated by a power law (7) with the average exponent $\alpha \approx$ 2.6. It should be noted, that dependence of these spectra on energy $E_\mu$ near the energy $E$ of the primary particle is very sharp. It is clearly seen for energies $3.162\times10^2$ GeV and $10^3$ GeV of the primary particles. The right parts of all curves when the energy $E_\mu$ of muons is approaching to a value of $E$ are very steep with the exponent $\alpha \approx 16$. It will be shown that these very steep parts that contribute substantially to the atmospheric muon flux.

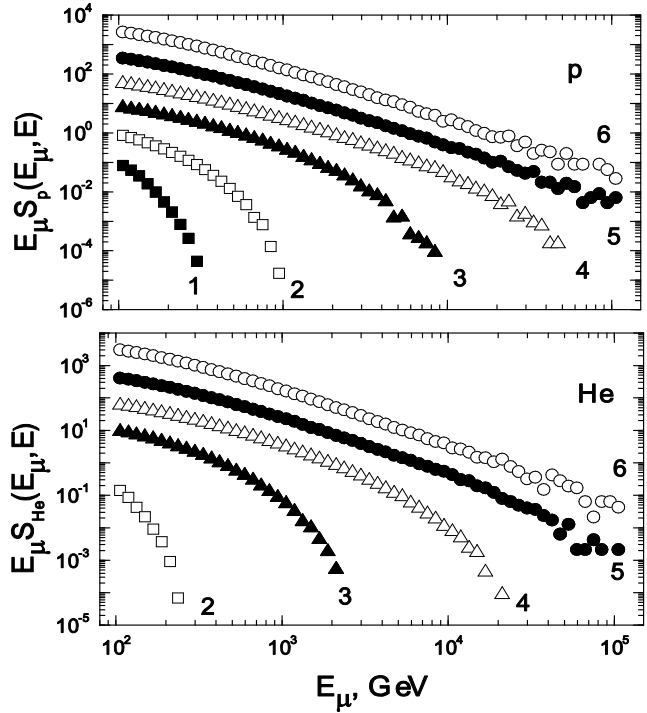

**Figure 2**. The distributions of muons $S_\mu(E_\mu,E)dE_\mu$ and $S_{He}(E_\mu,E)dE_\mu$ on the energy $E_\mu$ for the primary protons and helium nuclei with various fixed energies $E$ (1 – $3.162\times10^2$, 2 – $10^3$, 3 – $10^4$, 4 – $10^5$, 5 – $10^6$, 6 – $10^7$ GeV) for the model EPOS LHC [6].

To estimate this contribution the dependences of densities $S_p(E_\mu,E)$ and $S_{He}(E_\mu,E)$ on the energy $E$ of the primary particles should be displayed. As an example for the three bins with numbers $i = 1, 21, 41$ these dependences of the densities $S_p(\langle E_\mu(i)\rangle,E)$ and $S_{He}(\langle E_\mu(i)\rangle,E)$ on the energy $E$ of the primary particles calculated in terms of the EPOS LHC [6] model are shown in figure 3 for the the primary protons and helium nuclei. This figure displays first a sharp increase of these densities with growth of the energy $E$ and then power dependences

$$S_p(\langle E_\mu(i)\rangle,E) \sim E^\delta , \quad S_{He}(\langle E_\mu(i)\rangle,E) \sim E^\delta . \qquad (8)$$

The mean values of the exponent $\delta$ are equal to $0.85 \pm 0.05$, $0.85 \pm 0.05$ and $0.75 \pm 0.05$ for the models EPOS LHC [6], EPOS 1.99 [5] and QGSJET II-03 [2] respectively.

The range of the sharp increase takes nearly two orders of magnitude of the primary particle energy. This sharp increase corresponds to the steep drop of the densities $S_p(E_\mu,E)$ and $S_{He}(E_\mu,E)$ at the previous figure. It is this region that contribute substantially to the atmospheric muon flux. The "power" region is supressed due to the very steep energy spectrum of the primary particles. For these three bins mentioned above figure 4 demonstrates dependences of densities $S_p(\langle E_\mu(i)\rangle,E)$ and $S_{He}(\langle E_\mu(i)\rangle,E)$ multiplied by the "weight" functions $(dI_p/dE)$ and $(dI_{He}/dE)$ on the energy $E$ of the primary particles. These products which should be integrated on the energy $E$ of the primary particles for each bin of the energy distributions of muons due to the formulae (3). This figure demonstrates clearly the intervals of the energy $E$ of the primary particles which contribute substantially to the atmospheric muon flux. As an example this figure 4 demonstrates that for the bin $i$=1 the primary protons and helium nuclei contribute mainly at energies $E$ which are within the intervals $1.35 \times 10^2 - 1.44 \times 10^4$ GeV and $5.5 \times 10^2 - 7 \times 10^4$ GeV respectively. The maximal contributions happened to be at energies $6.54 \times 10^2$ GeV and $3 \times 10^3$ GeV of these primary particles respectively. As an energy per a nucleon is of importance these energy estimates for the the primary protons and helium nuclei are in agreement. It should be noted that at energies $E$ of the primary particles which are above $10^3$ GeV the number of fixed energies $E$ per one order of magnitude is two times less than at smaller energies. But figure 4 demonstrates that is practically unimportant.

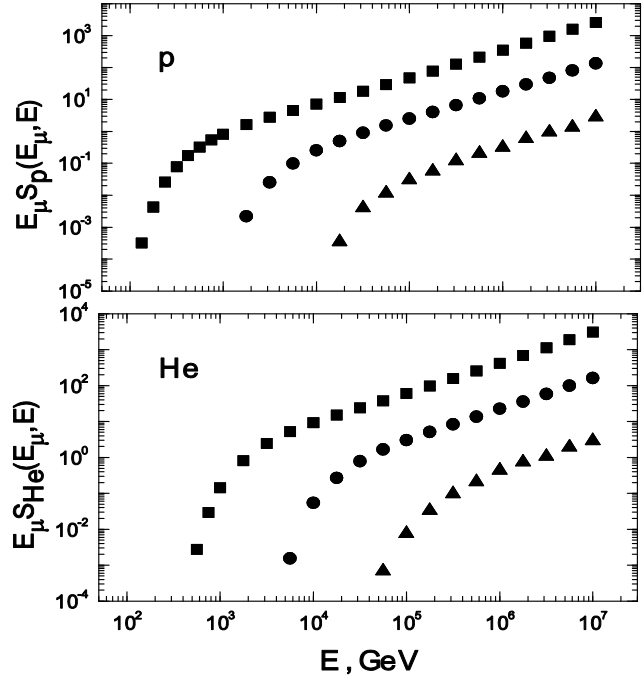

**Figure 3.** The distributions of muons $S_p(\langle E_\mu(i)\rangle, E)$ and $S_{He}(\langle E_\mu(i)\rangle, E)$ on the energy $E$ of the primary protons and helium nuclei for the model EPOS LHC [6] for bins $i$: ■ – 1, ● – 21, ▲ – 41.

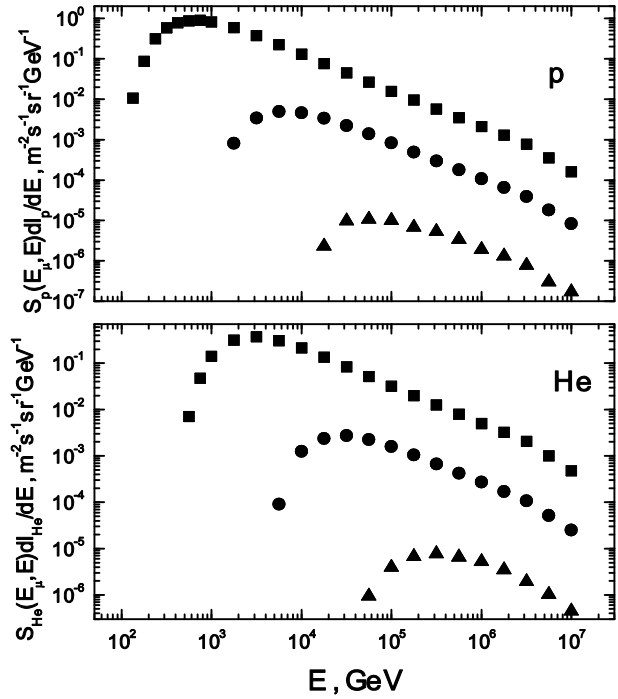

**Figure 4.** The same distributions as in figure 3 multiplied by the "weight" functions $(dI_p/dE)$ and $(dI_{He}/dE)$.

The dependence of densities $S_p(E_\mu, E)$ and $S_{He}(E_\mu, E)$ on the model used is also of interest. These muon densities calculated in terms of the various models (EPOS LHC [6], EPOS 1.99 [5] and QGSJET II-03 [2]) are shown in figure 5 for the primary protons and helium nuclei with the fixed energy $E = 10^5$ GeV. Results for the first two models are practically identical, but densities for last model are approximately for ~3 times and ~2.5 times smaller for the primary protons and helium nuclei respectively.

The spectra of the vertical atmospheric muons $D(E_\mu)dE_\mu$ in the energy range of $10^2 - 10^5$ GeV calculated in terms of EPOS LHC [6], EPOS 1.99 [5] and QGSJET II-03 [2] models are shown in figure 6. Only the primariy protons and helium nuclei were taken into account in our simulations. It can be seen that the EPOS 1.99 model [5] predicts the maximal intensity of the muon flux with highest energies. The EPOS LHC model [6] predicts a slightly lower flux.

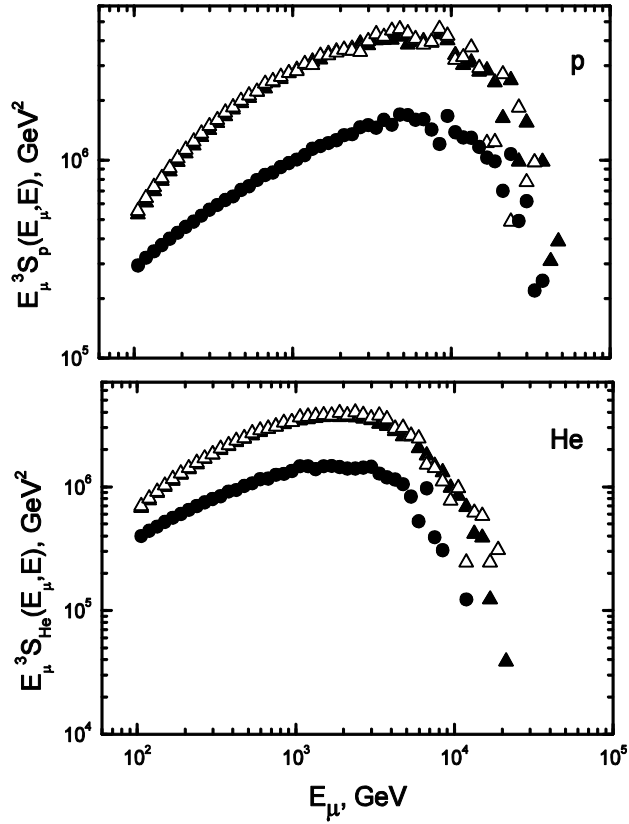

**Figure 5**. The distributions of muons on the energy $E_\mu$ calculated in terms of various models (▲ − EPOS LHC [6], ∆ − EPOS 1.99 [5], ● − QGSJET II-03 [2]) for the primary protons and helium nuclei with the fixed energy $E = 10^5$ GeV.

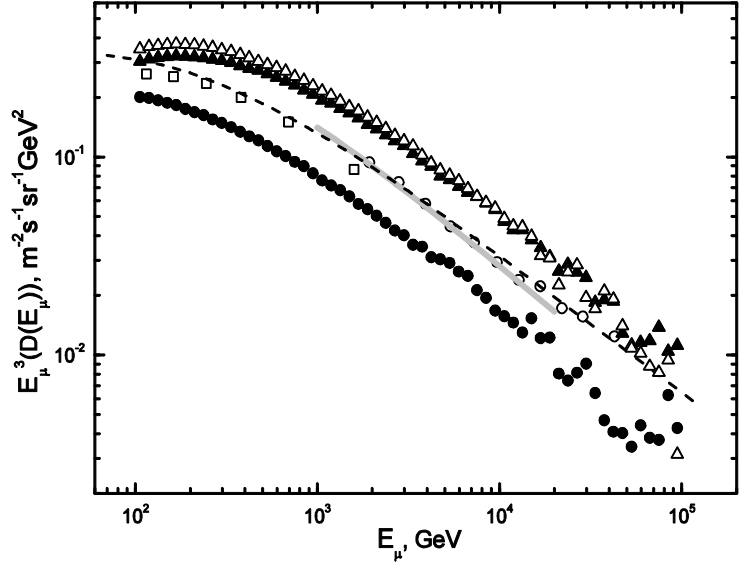

**Figure 6**. The energy spectra of the vertical atmospheric muons. ▲ − EPOS LHC [6], Δ − EPOS 1.99 [5] and ● − QGSJET II-03 [2] models; dashed line − the Gaisser approximation [41]; □ − L3+Cosmic [26], a grey stripe − MACRO [27], ○ − LVD [28].

The QGSJET II-03 [2] model gives considerably smaller flux. This conclusion is in agreement with results shown in figure 5. This figure shows also the Gaisser's approximation [41] of the muon energy spectrum and data observed by the collaborations L3+Cosmic [26], MACRO [27] and LVD [28]. Figure 6 also demonstrates the steepening of the spectrum at energies $E_\mu$ of muons which are much higher than the decay constant $B$ of $\pi^\pm$ mesons in the atmosphere ($B \approx$ 100 GeV). It is clearly seen that at energies of muons above $10^4$ GeV statistics of simulated events is not sufficient to make definite conclusions. So, a comparison of the calculated spectra with data allows testing models at highest energies of secondary particles.

Figure 7 demonstrates ratios $R$ of the energy spectra of muons simulated in terms of the EPOS LHC [6], EPOS 1.99 [5] and QGSJET II-03 [2] models to the smooth approximation of data observed by collaborations [26−28] for the energy interval $10^2$−$10^4$ GeV where uncertainties are not large. This figure shows that the ratios for the models EPOS LHC [6] and EPOS 1.99 [5] are monotonically increasing from the values of ~1.4 and 1.2 at the energy $10^2$ GeV up to 1.9 at $10^4$ GeV while these ratios for the QGSJET II-03 [2] model are decreasing from ~0.8 up to ~0.55 within the same energy interval. The most important fact is that this increase in ratios $R$ for the EPOS LHC and the EPOS 1.99 models exists already at energies $E_\mu$ ~$10^2$ GeV and becomes larger at $E_\mu$ = $10^4$ GeV. No slowing of this increase is observed at highest energies of muons. Thus, figure 7 demonstrates a very serious departure of the calculated spectra from data reported in [6−8]. This difference is associated with a slower rate of an energy fragmentation of projectile particles in their interactions with nuclei in the atmosphere. Thus, these models overestimate the probability of the generation of secondary particles with the highest energies. Contrary to these two models the QGSJET II-03 model predicts the muon flux at energy $10^4$ GeV which is by a factor ~1.8 smaller than data [26−28]. So, this model underestimates production of secondary particles with highest energies. It is a real reason that data [13] interpreted in terms of this model predicted heavy composition [14] at energies above $10^{17}$ GeV.

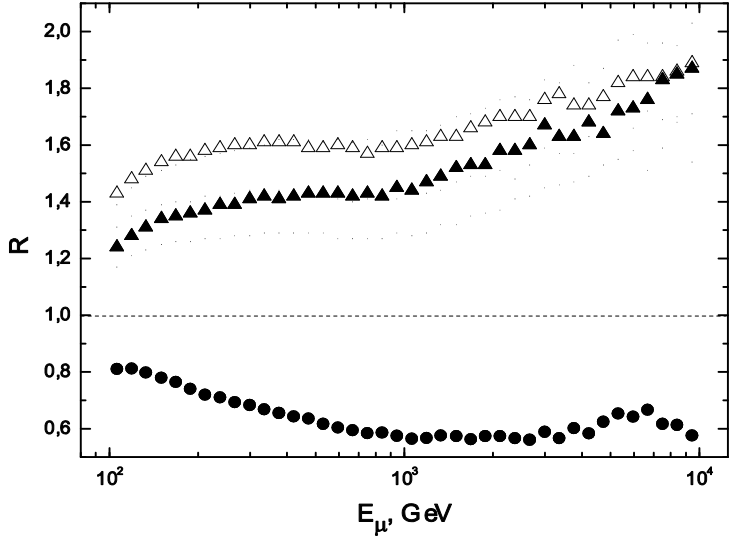

**Figure 7**. The ratios $R$ of the simulated vertical atmospheric muon fluxes to the smooth approximation of data [7−9]: ▲ − EPOS LHC [6], ∆ − EPOS 1.99 [5], ● − QGSJET II-03 [2] .

It should be noted that our results for the QGSJET II-03 [2] model are nearly 15−20% smaller than estimates calculated in [21] by the basically different approach with the ATIC-2 energy spectra [37] of the primary particles. Figure 8 shows the ratios $R$ of intensities of the muon energy spectrum $D(E_\mu)dE_\mu$ calculated in [21] in terms of the QGSJET II-03 model with the primary particle spectra [37] to results of our simulations in terms of the same model but for the GH spectrum [36] of the primary particles.

As the ATIC-2 energy spectra [37] is ~10% below at energy $10^2$ GeV and ~20% above the GH [36] approximation (see figure 1) this figure 8 reflects mainly the differences in the energy spectra of the primary particles observed in both collaborations. So, it is possible also to conclude that both approaches produce nearly the same results for the muon spectrum taking into account these differences in the energy spectra of the primary particles. Therefore, the accuracy of both methods is rather high.

This overestimated probability of secondary particle production at highest energies is also confirmed for the some models by the data of the LHCf [11] and TOTEM [12] accelerator experiments. For example, the QGSJET II-04 [3] overestimates the density of charged particles $dN_{ch}/d\eta$ per unit of pseudorapidity at the pseudorapidity $\eta = 6.345$ by a factor of $k \approx 1.3$ as compared to the TOTEM data [12]. This difference increases at larger $\eta$ values because of the difference in the slopes of the calculated curve and data from [12].

Comparison of the LHCf data [11] on the energy spectra of photons in *p-p* collisions at the energy of $\sqrt{s} = 7$ TeV with predictions of various models in the pseudorapidity range of $8.81 < \eta < 8.99$ shows that the QGSJET II-03 model [2] gives a two to four times smaller number of photons, whereas the EPOS 1.99 [5] model predicts a 1.5−2 times larger number of photons. Under the assumption of a similar dependence for charged mesons, this results in a decrease in the calculated density of muons at large distances from the axis of the shower. The immediate cause of this decrease odserved in [42, 43] is a displacement of a shower maximum to the deeper depth in the atmosphere due to a more slow rate of an energy fragmentation.

According to our calculations, the main contribution to integral (3) for each bin with the average energy $\langle E_\mu(i) \rangle$ comes from the primary protons and helium nuclei with energies in the ranges of $(1.3-1.4 \times 10^2) \cdot \langle E_\mu(i) \rangle$ and $(5.2-5.6 \times 10^2) \cdot \langle E_\mu(i) \rangle$, respectively. As these energy intervals are quite near the energy $E$ of a projectile particle only the first and probably the second generations of $\pi^\pm$ and $K^\pm$ mesons contribute to the muon flux in the atmosphere. If the first generation is overestimated by a factor $f$ then the second one will be overestimated by a factor $f^2$ and so on. So, the production of these secondary particles with the highest possible energies are overestimated in case EPOS 1.99 and EPOS LHC approximately by a factor of ~1.5 and underestimated by the same factor in case of QGSJET II-03. Thus, all models should be significantly corrected for the highest energies of secondary particles which are the most important for a development of EAS.

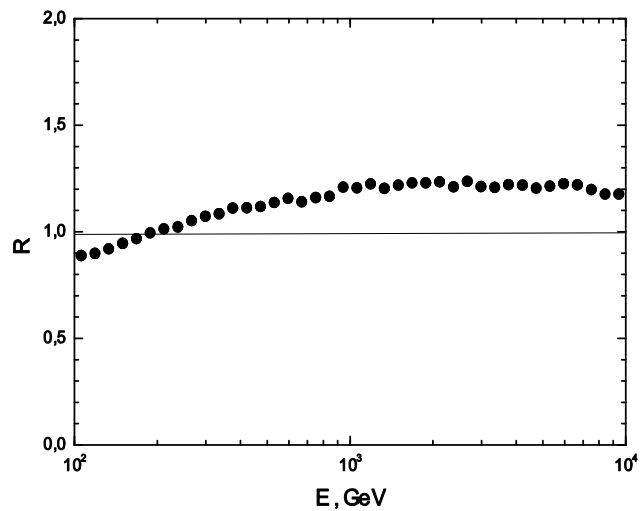

**Figure 8.** The ratios $R$ of intensities of the atmospheric muon flux [21] to our results.

**Acknowledgements**
Authors thank LSS (grant 3110.014.2) for support.